\documentclass[twocolumn,noshowpacs,preprintnumbers,secnumarabic,amsmath,amssymb,nofootinbib]{revtex4}

\usepackage{graphicx}
\usepackage{dcolumn}
\usepackage{bm}

\def\beq{\begin{equation}}
\def\eeq{\end{equation}}
\def\be{\begin{equation}}
\def\ee{\end{equation}}
\def\bea{\begin{eqnarray}}
\def\eea{\end{eqnarray}}
\def\d{{\rm d}}

\def\p{\partial}

\newcommand{\del}{\nabla}

\DeclareRobustCommand{\SkipTocEntry}[4]{}

\begin{document}


\vspace{5cm}
\title{On the Vacuum Propagation of Gravitational Waves}

\author{Xiao Liu}
\email{xliu@perimeterinstitute.ca}

\affiliation{%
Perimeter Institute for Theoretical Physics, Waterloo, Ontario,
Canada}

\date{May 31, 2007}

\begin{abstract}
We show that, for \emph{any} local, causal quantum field theory
which couples covariantly to gravity, and which admits Minkowski
spacetime vacuum(a) invariant under the inhomogeneous proper
orthochronous Lorentz group, plane gravitational waves propagating
in such Minkowski vacuum(a) do not dissipate energy or momentum via
quantum field theoretic effects.

\end{abstract}



\maketitle


\section{\label{sec:intro} Introduction}

Gravitational waves propagate in empty space in general relativity.
One basic class of such vacuum solutions is the plane waves in
Minkowski spacetime. They describe, for example, the propagation of
a gravitational wave, emitted by a bounded source, in a region far
from its source\footnote{For results of recent attempts of
experimental detection, see \cite{LigoFour}}. The metric is given by
the following exact solution to the vacuum Einstein equation

\beq \label{equ:Brinkman} \d s^2 = \d u \d v + h_{ij}(u) x^i x^j \d
u^2 - \d x^i \d x^i\, ,\eeq where $h_{ij}(u)$ is a symmetric,
traceless, and otherwise arbitrary $d-2$ by $d-2$ matrix valued
smooth function of $u$.  This describes a plane gravitational wave
propagating along the light-like direction $v$, with $h_{ij}(u)$
specifying the space-dependent profile for the $(d-2)(d-1)/2 - 1$
polarizations of the wave in $d$-dimensional spacetime\footnote{For
another coordinate system that makes explicit the symmetry of the
plane wave front, see e.g.{\cite{Blau}}}. $h_{ij}(u)$'s
in-dependence on the coordinate $v$ shows explicitly that, among
other things, no dissipation occurs. For $h_{ij} (u)$ with finite
support, the metric describes a plane gravitational wave in
Minkowski spacetime with a finite duration.

As classical solutions to the equations of motion, these plane wave
spacetimes are robust under a large class of deformations to the
gravitational dynamics. Any local correction to the Hilbert action
by adding higher powers of the Riemann tensor and covariant
derivatives leaves the solutions intact. This follows from the fact
that any scalar or second-rank tensor field constructed locally from
the Riemann tensor and the covariant derivative necessarily vanish
on these backgrounds\footnote{To jump ahead a little bit, we note
that this by itself does not imply that $\langle vac, in
|\hat{T}_{\mu \nu}^{ren}(x) |vac, in \rangle$ vanishes, which is, by
definition, sensitive to the \emph{global} property of the
background spacetime. A priori, non-vanishing second-rank tensor
fields can be constructed based on the Riemann tensor via
\emph{non-local} expressions.}. Hence any higher curvature and
higher derivative correction to the Hilbert action and to the
Einstein equation does not materialize.

In quantum theory, the vacuum is no longer empty. As ripples of
spacetime curvature travel through the vacuum, the zero-point
fluctuations of the quantum field in the background are generically
amplified to higher energy excitations. To calculate these effects,
one needs to diagonalize the full time-dependent Hamiltonian for the
quantum field at every instant of time, and to expand the in-vacuum
state in the basis of the out-states. This calculation is in general
difficult to carry out, except when the field theory is free. In the
latter case, diagonalizing the full Hamiltonian boils down to solving
free field equations in the time-dependent background, and the
occupation number of the in-vacuum in each out-Fock space state is
determined by the Bogolubov coefficients (see e.g.{\cite{birrell}}).

The metric ({\ref{equ:Brinkman}}) has a covariantly constant global
null Killing vector field, which generates the translation
\emph{along} the plane wave ($v$-direction). Had it generated an
evolution \emph{across} the wave (along $u$ for example), one would
have been able to immediately conclude that no particle would be
produced by the gravitational wave in \emph{any} field theory,
because one can choose to work with a (light cone)time-independent
Hamiltonian by slicing the geometry properly. Nevertheless, in the
case of \emph{free} quantum fields\cite{Gibbons}, the presence of
this along-the-wave Killing field is enough to forbid mixing between
positive and negative frequency modes under the evolution along the
$u$-direction across the wave (but still allows mixing between
positive/negative frequency modes with themselves respectively).
This establishes that plane gravitational waves do not dissipate
energy or momentum by exciting the vacuum of a free quantum field.

This calculation does not apply to interacting theories. In
particular, it does not apply to the real world, since the relevant
low energy physics is governed by the nonlinear interacting theory
of photons, in which the leading interaction from integrating out
the electron is $\frac{\alpha^2} {{m_e}^4} ((F^2)^2 + 7 (F
\widetilde{F})^2 )$\cite{Schwinger}. One may wonder whether, taking
into account this interacting nature of the QED (and the Standard
Model) vacuum, the gravitational wave could dissipate energy and
momentum by producing extremely soft photons as it
propagates\footnote{Such photons, if any, are presumably mostly
collinear to the gravitational wave on kinematic grounds.}.  Were
this possible to happen, attenuation of extremely high frequency
gravitational waves might accumulate over a cosmic distance scale to
a significant level \footnote{Dimensional analysis suggests that
such effects, even if present, are completely in-significant in the
frequency ranges of LIGO and LISA.}.

We analyze the physical vacuum expectation value of the renormalized
energy-momentum-stress tensor $\langle vac, in |\hat{T}_{\mu
\nu}^{ren}(x) |vac, in  \rangle$ in the plane gravitational wave
spacetime in general quantum field theories which couple to gravity
covariantly and which admit Minkowski space vacuum(a). This
expectation value is sensitive to the \emph{global} property of the
spacetime, and in principle can be computed from a regulated
$\langle vac, in | \hat{T}_{\mu \nu}(x)| vac, in  \rangle$ by
subtracting all cut-off dependence through
 \emph{local} counter-terms.\footnote{In addition to
counter-terms present in the flat space, new terms will be generated
in curved backgrounds involving the geometric quantities (the
Riemann tensor and covariant derivatives). Counter-terms that
involve only the geometric quantities vanish in the plane wave
background. Those that involve both the geometric quantities and the
other fields do not. In the case of interacting photon theory
coupled to gravity, for example, one such term that may be generated
and, if generated, does not vanish, is $\int d^d x  \sqrt{g}
R_{\alpha \beta \gamma \delta} F^{\alpha \beta} F^{\gamma \delta}$,
with a cut-off dependent coefficient. } Instead of pursuing this
direct approach case by case, we determine the form of the finite
piece $\langle vac, in |\hat{T}_{\mu \nu}^{ren}(x)|vac, in \rangle$
for a very general class of field theories by exploiting the
symmetries of the plane wave spacetime. We find that $\langle vac,
in |\hat{T}_{\mu \nu}^{ren}(x) |vac, in \rangle$ vanishes
identically in the metric ({\ref{equ:Brinkman}}). This conclusion
holds \emph{regardless} of the nature and the strength of the
interaction\footnote{The interaction, for example, does \emph{not}
need to preserve $P$, $T$, $C$ separately in Minkowski spacetime.}
and the phase of the vacuum, and independent of the dimension of the
spacetime. It shows that gravitational waves far from its source
propagate without dissipation via any quantum field theoretic
effects.

\section{\label{sec: geometry} Gravitational Plane Wave Spacetime}

The spacetime defined by the metric (\ref{equ:Brinkman}) is
geodesically complete, and contains no closed time-like or
light-like curves. It admits $2d-3$ Killing vector fields for generic
choice of $h_{ij}(u)$, although only one of them

\beq \label{equ:null} Z = \frac{\p}{\p v}\, ,\eeq is manifest in the
Brinkman coordinates (\ref{equ:Brinkman}).
The $2(d-2)$ non-manifest
Killing vector fields are all in the form:

\beq\label{equ:killing} X = 2 b_i'(u) x^i \p_v + b_i(u)\p_i \, ,\eeq
where $(b_1(u),\,.\,.\,.\,,b_{d-2}(u))$  is a solution to:

\beq\label{equ:harm} b_i''(u)+ h_{ij}(u)b_j(u)=0 \, .\eeq The ODE
has $2(d-2)$ independent solutions, which give rise to the same
number of additional independent Killing vector fields. Killing
fields associated to two solutions $b_i(u)$ and
$\tilde{b}_j (u)$ satisfy

\beq\label{equ:comm} \left[X_{b}, X_{\tilde{b}}\right]= 2 W [b, \tilde{b}] Z\,
,\eeq where the Wronsky $W[b, \tilde{b}]= \sum_i(b_i(u) \tilde{b}_i' (u) - b_i' (u)\tilde{b}_i(u) )$
is independent of $u$.
In a suitable basis, they generate the Heisenberg algebra 

\beq\label{equ:heisen} [X_{(k)}, \widetilde{X}_{(l)}]=\delta_{kl} Z
\,\,\,\,\,\,\,\,\{k,l=1,...,d-2\}\, ,\eeq with central element Z.

The Killing vector fields in (\ref{equ:heisen}) preserve each $u=$const hypersurface, and generate
on each such hypersurface  the $d-1$ translations and the $d-2$ $x$-linearly-dependent translations
along $v$. For any given Killing vector field, the actions on the constant $u$ hypersurfaces
are $u$-dependent. To help characterize the algebraic aspect of this dependence,
we introduce yet another vector field

\beq\label{equ:hamil} H = \frac{\p}{\p u}  \, .\eeq This is
\emph{not} a Killing field. It generates evolution along $u$ and
would be upgraded into the light-cone Hamiltonian if we quantize
field theories in the plane wave background. In the basis of
(\ref{equ:heisen})

\bea \left[H, X_{(k)}\right] &=& -h_{kl}(u) \widetilde{X}_{(l)}
\nonumber\, ,\\ \left[ H, \widetilde{X}_{(l)} \right] &=& X_{(l)}
\label{equ:evol}\, ,\eea and $Z$ remains central. Observe that the
algebra does \emph{not} close, unless $h_{ij}(u)$'s are constant.
This is what one expects, since $H$ does not generate isometries
unless the metric does not depend on $u$.

We are interested in plane gravitational waves of finite duration,
that propagate in otherwise flat spacetime. So we demand $h_{ij}(u)$
to vanish outside $[-T, T]$. In regions $| u | >T $ where
(\ref{equ:Brinkman}) reduces to flat space, the Killing fields
generate $a$ subgroup of $the$ subgroup of Poincare group that
preserves the null hyperplanes $\{u=u_0\}$. The latter
is the same subgroup that preserves the vector field $\frac{\p}{\p v}$, and
is generated by the translation along $v$, the translations and rotations
of the $x^i$ 's among themselves, and the $d-2$ additional vector fields:

\beq\label{equ:boosrot} 2 x^i \p_v + u \p_i \, .\eeq The last $d-2$  fields are
linear combinations of boosts and rotations
. All the
translations and the $d-2$ boost-rotations extend to the
whole plane wave spacetime, while the rotations among the $x^i$
themselves do not for generic $h_{ij}(u)$. The translations and
boost-rotations account for the total $2d-3$ global Killing vector fields.

\section{Plane Waves as Robust CLASSICAL Solutions in effective field theories}

The Einstein-Hilbert action is an effective action for gravity. Various
higher dimensional operators may be added and are presumably indeed present,
their effects small until spacetime curvature approaches the mass scale that
suppresses these higher dimensional operators, at which point the applicability
of the effective theory itself starts to break down.
Two natural questions come to mind: (1)what modifications these corrections may bring to the
gravitational wave solutions? (2)how strong the gravitational wave needs to be
to invalidate the application of the effective theory itself? In normal  
scenarios, the cut-off scale for the gravity effective action is assumed to be not too
low below the Planck scale or the string scale.

The answer to the first question is well-known (see, e.g.
\cite{Deser} \cite{Gary}): any higher curvature and higher derivative corrections
to the Einstein-Hilbert action, involving only the quantities
derived from the metric itself, does not modify the geometry
({\ref{equ:Brinkman}}).

The reason for this, as already mentioned in the introduction, is
that any scalar and non-trivial second-rank tensor field that can be
constructed based on the metric, the curvature, and the covariant
derivative, necessarily vanish in the background
({\ref{equ:Brinkman}}).  Hence both the corrections to the action,
and the corrections to the Einstein equation, vanish for the plane
wave spacetime.

To see how the geometric property comes about, we compute the
Riemann tensor and its covariant derivatives in the Brinkman
coordinate basis. The only nonvanishing components of the Riemann
tensor are

\beq\label{riemann} R_{uiuj}=-h_{ij}(u) \, ,\eeq and those related
to this by symmetry. Further inspection reveals that
$\del_{\alpha}\del_{\beta}... \del_{\gamma} R_{\mu \nu \rho \sigma}$
vanishes unless every index is either $u$ or one of the $d-2$ $i$'s,
\emph{and} the total number of $i$-index must be less than or equal
to $2$. In fact, by inspecting the basic operations in the
construction of these higher rank tensor fields, a simple ``sum
rule'' can be shown to hold: the total number of $i$-index for any
nonvanishing component plus its degree as a homogeneous polynomial
of the variables $\{x^i, i=1,...,d-2\}$ always equals to $2$.
Technicalities aside, the upshot for now is that any component of
the above tensors with a $v$-index vanishes. Note also that $g^{u
\nu} \neq 0$ if and only if $\nu$ is $v$, and that $h_{ij}(u)$ is
traceless. It then follows by inspection that no non-vanishing
scalar or second-rank tensor fields (except the metric itself) can
be constructed because there are too many lower $u$ indices and no
lower $v$ index that there is no way to contract all of them.

Since the contribution of all higher dimensional operators are
parametrically smaller (indeed, in this case, they vanish) than the
leading contribution from the Einstein equation,  the application of
the effective theory is, \emph{by definition}, valid regardless of
how strong the gravitational wave is. This might seem a little
confusing at the first glance, because the solutions
({\ref{equ:Brinkman}}) allow $h_{ij}(u)$, the \emph{tidal force},
arbitrarily large and arbitrarily fast varying.

There is no paradox here. The point is that $h_{ij}(u)$ and its variations are
frame-dependent; and, around every point in the plane wave geometry, one
can always boost to a free-falling frame in which
all components of the tidal force and its
gradients are smaller than 1 in \emph{any} specified mass unit.
So they can all be made parametrically small compared to
the mass scales suppressing the corrections in the action.
This follows from the non-compactness of the Lorentz group and the
the fact above that there is no non-vanishing
Lorentzian scalar constructed locally by the Riemann tensor and its
derivatives.
To gain intuitions about it, we now see it directly from the
metric ({\ref{equ:Brinkman}}).

Along the locus \{$x^i=0$,\, $i=1,...,d-2$\}, the basis
$\{\frac{1}{2}(\p_u + \p_v),\, \frac{1}{2}(\p_u - \p_v),\, \p_{i},\,
i=1,...,d-2 \}$ is already a Lorentz frame at each point, namely the
metric tensor in this basis is $diag(+1, -1,...,-1)$. A boost $\{
\p_u\mapsto \lambda \p_u$,\,$\p_v\mapsto\lambda^{-1} \p_v$,\,
$\p_{i} \mapsto \p_{i} \}$ (regarded as a linear transformation in
the tangent space at a given point) leaves the metric invariant, but
enforces $\del_{\alpha_1}...\del_{\alpha_k} R_{\mu \nu \rho \sigma}
\mapsto \lambda^{k+2} \del_{\alpha_1}...\del_{\alpha_k} R_{\mu \nu
\rho \sigma}$ for every nonvanishing component, by the previously
mentioned sum rule. Hence given any finite number of such tensors,
we can always choose $\lambda$ properly to make all the components
of these tensors arbitrarily small. This finds the proper Lorentz
frames point-wise along $\{x^i=0,\,\, i=1,...,d-2\}$.

Remember that the spacetime has a large isometry, which acts on
each hypersurface $\{u=u_0\}$ transitively, some of which, specifically,
act as translations. So for any point $P$ in the spacetime, there
always is an isometry that brings that point to a point $Q$ in $\{x^i=0,\,\,
i=1,...,d-2\}$. The pullback to $P$ from $Q$ of the appropriate
Lorentz frame at $Q$ gives rise to the sought-after frame at $P$ that makes
all the components of the tidal force and its gradients small.

The boost we did to scale down the tidal force corresponds to
speeding up in the direction that the wave propagates. This
elongates the duration of the plane wave and lowers its frequency of
variation, both of which are frame-dependent scales. Furthermore, as
shown above, no frame-independent scale exists at all that
\emph{local} observer can define in the spacetime\footnote{To jump
ahead a little bit, one may try to infer the absence of particle
production based on an adiabatic reasoning. The author does not know
how, along this line of reasoning, to rule out extensive production
effects in cases without a mass gap. What we will do instead is to
determine the effects by exploiting its property under boosts and
other transformations.}. This implies that the solutions
(\ref{equ:Brinkman}) are valid for arbitrarily strong waves. On the
other hand, as well-known, the field theoretic description does
break down, but only as we start asking questions about the local
physics on length scales approaching the cut-off scale.

To recapitulate, as long as we restrict ourselves to length scales above the cut-off,
not only that the geometry ({\ref{equ:Brinkman}}) is always valid as classical
solutions to the effective action, but also that
the application of the effective action itself is
always valid in solving for these classical solutions.

\section{\label{sec: calculation}
$\langle vac, in | \hat{T}_{\mu \nu}^{ren}(x) |vac, in  \rangle $ in
Generally Covariant Quantum Field Theories}

We ignored the presence of other fields in the last section by
setting them to their values in a classical vacuum, which we assume
to be a configuration of Minkowski space with a Poincare invariant
profile for all the fields present. This is consistent at the
\emph{classical} level \footnote{ That the other fields do not
backreact on spacetime is clear because $T_{\mu \nu}$ vanishes in
the classical vacuum. That the gravitational wave does not disturb
the fields away from their vacuum values requires some
qualification. We assume that covariant couplings to gravity linear
in the fields vanish. This is automatically satisfied in the plane
wave spacetime for particles with spin less or equal to 3/2. In the
presence of higher spin particles, we would have to impose this
extra assumption.}.

Quantum mechanically, local observables in the vacuum are no longer
sharply peaked at any particular values. The consequences are
several-fold. First, zero-point motions give rise to cut-off
dependent contributions to the effective Lagrangian density. These
include, generically, a constant piece, acting effectively like a
cosmological constant, and various other fields and curvature
dependent terms. Unless one works in a UV finite theory like string
theory, one can only determine the coefficients of these
interactions by measurements. We will make no statements about these
coefficients, except that we restrict ourselves to theories that
admit (proper orthochronous) Poincare invariant Minkowski space
vacuum(a). This implies, in particular, that the total cosmological
constant vanishes, and that only operators which transform as
(pseudo-)scalars may acquire vacuum expectation values.

Suppose now that we have experimentally determined all the couplings of the
interactions in the effective theory and computed the ground state wave function of the
quantum fields in the Minkowski vacuum. We ask the question, what happens to the quantum
field vacuum as a train of plane gravitational wave passes by. As mentioned
in the introduction, one $a$ $priori$ expects that particles would be excited, although
a calculation specific to free fields suggests otherwise.

So we consider quantizing a general field theory \footnote{We
exclude gravitons from this field theory, because the
energy-momentum tensor of gravitons, constructed from $h_{\mu \nu}$,
is not a tensor of the Lorentz group.}in the background
({\ref{equ:Brinkman}}) for which $h_{ij}(u)$ has a finite support
$[-T, T]$. We specify the initial condition that at $\{u=u_0< -T\}$,
the theory lives in the in-vacuum $|vac, in\rangle$. To fully
specify the system, we also need to impose boundary conditions along
$\{u > u_0,\,\,v \rightarrow -\infty\}$. We require that the
boundary condition at $v \rightarrow -\infty$ preserves the isometry
of the spacetime. This implies, in particular, that no field and/or
particle comes in from $v \rightarrow -\infty$ except the
gravitational wave itself.

We analyze the conditions $\langle vac, in | \hat{T}_{\mu
\nu}^{ren}(x) |vac, in  \rangle $ needs to satisfy. Remember that in
the Minkowski space portion of the spacetime ({\ref{equ:Brinkman}}),
the isometries are part of the Poincare group which, by assumption,
leaves $|vac, in\rangle$ invariant. Combined with the boundary
condition we imposed, this implies that $\langle vac, in |
\hat{T}_{\mu \nu}^{ren}(x) |vac, in  \rangle $ (for convenience, we
will denote this quantity by $\langle T_{\mu \nu} \rangle$ ) is an
invariant tensor under the isometry group. That is, for any
transformation $x \mapsto y=f(x)$ that satisfies $(f^* g)_{\mu \nu}=
g_{\mu \nu}$ we have \beq\label{equ:defiso} \frac{\p y^{\rho}}{\p
x^{\mu}} \frac{\p y^{\sigma}}{\p x^{\nu}} \langle T_{\rho \sigma}(y)
\rangle = \langle T_{\mu \nu}(x) \rangle\, .\eeq

That this does not only hold in the before-wave region but also hold everywhere
requires some explanation. Let $\hat{U}[f]$ be the operator
that realizes the isometry transformation $f$ in the quantum field
theory. This operator is $u$-dependent, and
its action on each constant $u$ hypersurface, which the isometry $f$
preserves, is determined by the generating Killing vector field
$V_f$, which, in turn, is determined by (\ref{equ:killing}) or
equivalently by ({\ref{equ:evol}}). We have \beq\label{equ:tens}
\frac{\p y^{\rho}}{\p x^{\mu}} \frac{\p y^{\sigma}}{\p x^{\nu}}
\hat{T}_{\rho \sigma}^{ren}(y)  = \hat{U}[f]^\dag(u)\,\cdot
\hat{T}_{\mu \nu}^{ren}(x)\, \cdot \hat{U}[f](u)\, \eeq by the fact
that $\hat{T}_{\mu \nu}^{ren}(x)$ is an operator that transforms as
a tensor; the arguments of $\hat{T}$ in this equation (points $x$
and $y$) share the same value for the $u$-coordinate.

We claim that, for \emph{all} values of $u$, $\hat{U}[f](u)$ leaves
the in-vacuum invariant. This is clear if $|u|\geq T$ in which case
it represents an element of the Poincare group that preserves the
null hyperplanes $u=$constant. It might seem less clear if $|u|<T$,
but it is also true. The point is, on each constant $u$
hypersurface, the Killing vector field $V_f$  that generates $f$ can
be expanded in the basis of vector fields $\{\p_v,\, \p_i,\, x^j
\p_v,\,\,\,i,j=1,.\,.\,.\,,d-2\}$ restricted to the same
hypersurface. The corresponding operator $\hat{O}[V_f] (u)$ can thus
be expanded in terms of the $u$-independent operators
$\{\hat{O}[\p_v],\, \hat{O}[\p_i],\, \hat{O}[x^j \p_v],\,\,\,
i,j=1,.\,.\,.\,,d-2\}$ with $u$-dependent coefficients. Since we
know from the flat before-wave region that all the latter annihilate
the in-vacuum, $\hat{O}[V_f](u)$ must also do. Hence,
$\hat{U}[f](u)$ leaves it invariant:

\beq\label{equ:inva} \hat{U}[f](u)\,\, |vac, in \rangle =  |vac, in
\rangle\, \eeq for all values of $u$. Now we sandwich
({\ref{equ:tens}}) between $\langle vac, in |$ and $|vac, in
\rangle$, simplify the right hand side via
({\ref{equ:inva}}), and produce ({\ref{equ:defiso}}).

It follows from ({\ref{equ:defiso}}) that

\beq \label{equ:iso} \pounds_V  \langle T_{\mu \nu} \rangle = 0\,
,\eeq for any $V$ in the algebra ({\ref{equ:heisen}}). Writing out
this equation explicitly we find that for $V=Z$

\beq\label{equ:z} \p_v \langle T_{\mu \nu} \rangle=0 \, \eeq and that for $V$ one of the
$X$'s

\bea 0 &=& b_i(u)(\p_i \langle T_{\mu \nu} \rangle- 2
h_{ij}(u) x^j (\delta_\mu^u \langle T_{v \nu} \rangle) + \delta_\nu^u \langle T_{\mu v} \rangle)
\nonumber\\
&+& b_i'(u) \langle T_{\mu \rho} \rangle (2 \delta^\rho_v \delta_\nu^i + \delta^\rho_i
\delta_\nu^u)\nonumber\\
&+& b_i'(u) \langle T_{\rho \nu} \rangle (2 \delta^\rho_v \delta_\mu^i
 + \delta^\rho_i \delta_\mu^u)\, \label{equ:X}\eea where, as before, $i,j=1,...,d-2$
and repeated indices are summed over regardless of their vertical
positions. Since the $2d-4$
$(b_1(u),\,.\,.\,.\,,b_{d-2}(u))$'s that define the $X$-type isometries
constitute a complete basis of the
solutions to (\ref{equ:harm}), the functions multiplying $b_i(u)$ and
$b_i'(u)$ in (\ref{equ:X}) must vanish separately. Working out their consequences
we find

\bea \langle T_{vv} \rangle &=& \langle T_{vi} \rangle =  \langle T_{ui} \rangle = 0\nonumber\, ,\\
 \langle T_{uu} \rangle (u, x^i) &=& 2 h_{ij}(u)x^i x^j \langle T_{uv} \rangle
(u) \, ,\label{equ:result} \\ \langle T_{ij}\rangle (u) &=& -2  \langle T_{uv}
\rangle (u) \delta_{ij }\nonumber\, .\eea
In the above
equations, we explicitly write out the coordinate(s) each component
of $\langle T_{\mu \nu}\rangle$ is allowed to depend on; for
example, $\langle T_{uv} \rangle$ can only depend on $u$. These all
follow from solving the isometry constraints ({\ref{equ:iso}}).

It is now clear that, that $\langle vac, in | \hat{T}_{\mu
\nu}^{ren}(x) |vac, in  \rangle $ is invariant under the full
isometry group is very constraining: the expectation values of a
$d$-dimensional second-rank symmetric tensor, that is $d(d+1)/2$
functions of $d$ variables each, reduce to a single unknown function
$\langle T_{uv} \rangle$ of a single variable $u$! This is certainly
only possible because we started in the in-vacuum and imposed proper
boundary conditions, any excitations in the initial state or in the
in-coming wave from $v \rightarrow -\infty$ will spoil the property.

To proceed further, we will need some dynamical equation, which is
generally hard to write down. There is a simple one, the covariant
conservation of the energy-momentum-stress tensor:

\beq\label{equ:conserv} \del^{\mu} \langle T_{\mu \nu} \rangle = 0\,
.\eeq This condition is necessary for general covariance to be
preserved at the quantum mechanical level{\cite{witten}}. Now
applying the results in ({\ref{equ:result}}), it simplifies to

\beq\label{equ:semifinal} 2 \p_u \langle T_{v \nu}\rangle + \p_i \langle T_{i \nu} \rangle = 0 \, .\eeq

Further application of ({\ref{equ:result}}) immediately shows that
(\ref{equ:semifinal}) only gives nontrivial constraint when $\nu$
is $u$:

\beq\label{equ:final} \frac{\d} {\d u} \langle T_{uv} \rangle (u) =0
\, .\eeq Hence $\langle T_{uv} \rangle$ is a constant.

What we have shown is that, up to an overall constant, there is
precisely one covariantly constant symmetric second rank tensor
field in the background ({\ref{equ:Brinkman}})that is invariant
under the full isometry (\ref{equ:heisen}). Of course, the metric
tensor itself satisfies these conditions, hence $\langle T_{\mu \nu}
\rangle = constant \times g_{\mu \nu}$. Since we started from a
Minkowski vacuum in which $\langle T_{\mu \nu} \rangle \equiv 0 $
for $u< -T$, this constant must vanish \footnote{We thank K.Krasnov
for pointing out the reference {\cite{UV}} which showed that the
cosmological constant, if zero, is not renormalized by pure graviton
loops up to two loops. If such result fails to hold at higher loops
and/or after coupling with matter, $\langle T_{\mu \nu} \rangle =
constant \times g_{\mu \nu}$ by itself means that no dissipation of
energy and momentum into the matter sector occurs.} .

We showed that $\langle vac, in | \hat{T}_{\mu \nu}^{ren}(x) |vac,
in  \rangle \equiv 0$ in the gravitational plane wave background.
What does it mean? Had the in-vacuum evolved into an (locally
discernable) excited out-state in the future flat region, this
quantity would have been non-vanishing. Hence, for any local
observer after the wave, the field appears to remain in a vacuum
state. Put another way, in \emph{any} finite (however large) region
of space, the energy and momentum dissipated into the quantum field
in that region by the gravitational wave vanish exactly.

On the other hand, the \emph{gravitational aspect} of $\langle
\hat{T}_{\mu \nu}^{ren} \rangle$'s significance is not at all clear
at the conceptual level. Plausible statements had been made in the
literature that suggest to feed it back to the Einstein equation to
further correct the background metric in some sort of semiclassical
approximation, but none had been made precise. Time-dependent
backgrounds in string theory would hopefully be understood
well-enough to clarify its physical significance in future.
Nevertheless, we note, given that the field theoretic aspects of the
computation of $\langle \hat{T}_{\mu \nu}^{ren} \rangle$ is
well-defined, the result $\langle \hat{T}_{\mu \nu}^{ren} \rangle
=0$ should be taken seriously. It may also be reassuring to note
that, incidentally, this result nullifies further concerns of
backreaction on the metric at the semi-classical level.

\section{Remarks}

When solving the free field wave equation in the plane gravitational
wave spacetime {\cite{Gibbons}}, one finds that a monochromatic
positive frequency solution in the before-wave region evolves into a
\emph{superposition} of positive frequency solutions after the wave
passes by\footnote{To see this, one needs to transform the equations
(3.1)-(3.3) of {\cite{Gibbons}}, which are given in the Rosen
coordinate associated to the before-wave region, into the global
Brinkman coordinate. One should also note that the singularities of
the mode solutions do \emph{not} represent a fundamental obstruction
to quantizing the field theory. They disappear when wave packets are
considered that have \emph{finite} supports in directions transverse
to the propagation of the wave. On the other hand, it does indicate
formations of singularity when two infinite plane waves are
collided{\cite{Yurtsever}}.}. That is, the creation operators do not
mix with annihilation operators, but they do mix with themselves. In
a free field theory, for which the physical ground state is the same
as the Fock space ground state, one concludes that the field stays
in the vacuum undisturbed. After interaction is turned on, one
expects that the physical ground state spreads out in the Fock
space. So it may appear that mixing between the positive frequency
solutions themselves would generically lead to volume-extensive
particle productions. We find the contrary.

The point has to do with the vacuum structure in light cone
quantization. Remember that we sliced the geometry by constant $u$
hypersurfaces, which are light-like. The rules of light cone
quantization for general field theories are not entirely clear, but
it is generically expected that the physical vacuum, modulo the
zero-modes problem, is the same as the Fock space
vacuum{\cite{burkardt}}, as a result of the positivity of the
longitudinal light cone momentum. This vacuum furthermore is not
affected when the light cone Hamiltonian becomes (light cone)
time-dependent, again modulo the problem associated to the
zero-modes. In a simple case like the $\lambda \phi^4$ theory, the
light cone quantization in the plane wave spacetime can be carried
out and explicitly shows that no particle production effects arise.
On the other hand, our argument in the previous section holds for
general field theories. It does not depend on any specifics of light
cone quantization, but is consistent with expectations derived from
it.


\medskip

The message to take away is that, the propagation of gravitational
waves in stable (proper orthochronous) Poincare invariant Minkowski
spacetime vacua is robustly characterized by the classical solutions
to general relativity. \emph{Vacuum fluctuations of field theoretic
origin, regardless of their property, do not modify this behavior}.

\smallskip

Exceptions, however, may arise if Lorentz invariance is
spontaneously broken in a Minkowski vacuum. One such class of
examples was constructed, at low energies, in {\cite{Nima}}. After
coupled to gravity, the goldstone field $\pi(x)$ of these theories
violates the equivalence principle and allows sources to
anti-gravitate; $\pi(x)$ also develops a Jeans-like instability at
large distances around the flat background. It may be interesting to
study the propagation of gravitational waves in this class of
Lorentz violating vacua, both at the classical and at the quantum
level.

\smallskip

For practical purposes, it is perhaps important to study the
propagation of a gravitational wave in a gas of particles ( see e.g.
{\cite{bondi1} \cite{bondi2} \cite{bondi3}} for some earlier
results).


\addtocontents{toc}{\SkipTocEntry}
\section*{Acknowledgments}
The author would like to thank Robert Brandenberger, Jaume Gomis,
Gary Horowitz, Shamit Kachru, Justin Khoury, Slava Mukhanov, and
Constantinos Skordis for helpful comments on the draft. He would
also like to thank Freddy Cachazo and Amihay Hanany for suggestions
of references on light cone quantization. Research at the Perimeter
Institute for Theoretical Physics is supported in part by the
Government of Canada through NSERC and by the Province of Ontario
through MRI.



\begingroup\raggedright\endgroup

\end{document}